\documentclass[10pt, conference, compsocconf]{IEEEtran}

\usepackage{algorithm}
\usepackage{algorithmicx} % Doc is at http://tug.ctan.org/macros/latex/contrib/algorithmicx/algorithmicx.pdf
\usepackage{algpseudocode}
\usepackage{amsfonts} % for R symbol (the set of real numbers)
\usepackage{color}
\usepackage{colortbl} % for \rowcolor
\usepackage[pdftex]{graphicx}
\usepackage{graphicx}
\usepackage[bookmarks=false]{hyperref}
\hypersetup{colorlinks=true,linkcolor=black,citecolor=black,filecolor=black,urlcolor=blue}
\usepackage{mathtools}
\usepackage{multirow}
\usepackage{stmaryrd} % for llbracket and rrbracket
\usepackage{subcaption}
\usepackage{nicefrac}
\usepackage{amsmath}
\usepackage{amssymb}
\usepackage{stmaryrd}
\usepackage{amsthm}
\usepackage{wrapfig}
\usepackage{multicol}
\usepackage{flushend}

\newlength{\bibitemsep}
\newlength{\bibparskip}
\let\oldthebibliography\thebibliography
\renewcommand\thebibliography[1]{%
  \oldthebibliography{#1}%
  \setlength{\parskip}{\bibitemsep}%
  \setlength{\itemsep}{\bibparskip}%
}

\algblock{Input}{EndInput}
\algnotext{EndInput}
\algblock{Output}{EndOutput}
\algnotext{EndOutput}
\newcommand{\Desc}[2]{\State \makebox[2em][l]{#1}#2}

\newcommand{\norm}[1]{\left\lVert#1\right\rVert}
\newtheorem*{theorem}{Theorem}
\definecolor{headcolor}{gray}{0.9}

\begin{document}

\title{A multi-dimensional extension of the Lightweight Temporal Compression method}

\author{Bo Li$^1$, Omid Sarbishei$^2$, Hosein Nourani$^1$, Tristan Glatard$^1$\\
  $^1$ Department of Computer Science and Software Engineering, Concordia University, Montreal, QC, Canada \\
  $^2$  Research and Development Department, Motsai Research, Saint Bruno, QC, Canada \vspace*{0.9cm}}
\maketitle

\begin{abstract}
Lightweight Temporal Compression
(LTC) is among the lossy stream
compression methods that provide the highest compression rate for the 
lowest CPU and memory consumption. As such, it is well suited to 
compress data streams in energy-constrained systems such as connected 
objects. The current formulation of LTC, however, is one-dimensional 
while data acquired in connected objects is often multi-dimensional: for instance, 
accelerometers and gyroscopes usually measure variables along 
3 directions. In this 
paper, we investigate the extension of LTC to 
higher dimensions. First, we provide a formulation 
of the algorithm in an arbitrary vectorial space of dimension $n$. 
Then, we implement the algorithm for the infinity and Euclidean norms, 
in spaces of dimension 2D+t and 3D+t. We evaluate our implementation on 
3D acceleration streams of human activities. 
Results show that the 3D implementation of LTC can save up to 20\% in 
energy consumption for low-paced activities, with a memory usage of about 100~B.
\end{abstract}

\section{Introduction}

With the recent technological advances in Internet of Things (IoT) 
applications, more than one billion connected objects are expected to 
be launched worldwide by 2025\footnote{\url{ 
https://www.statista.com/statistics/471264/iot-number-of-connected-devices-worldwide}}.
Power consumption is among the biggest challenges targeting connected 
objects, particularly in the industrial domains, where several sensing 
systems are commonly launched in the field to run for days or even 
weeks without being recharged. Typically, such devices use sensors to 
capture properties such as temperature or motion, and stream them to a 
host system over a radio transmission protocol such as Bluetooth 
Low-Energy (BLE). System designers aim to reduce the rate of data 
transmission as much as possible, as radio transmission is a 
power-hungry operation.

 Compression is a key technique to reduce the rate of radio 
 transmission.  While in several applications lossless compression 
 methods are more desirable than lossy compression techniques, in the 
 context of IoT and sensor data streams, the measured sensor data 
 intrinsically involves noise and measurement errors, which can 
 be treated as a configurable 
tolerance for a lossy compression algorithm. 

Resource-intensive lossy compression algorithms such as the ones based on 
polynomial interpolation, discrete cosine and Fourier transforms, or 
auto-regression methods~\cite{lu2010optimized} are not well-suited for 
connected objects, due to the limited memory available on 
these systems (typically a few KB), and the energy consumption 
associated with CPU usage. Instead, compression algorithms need 
to find a trade-off between reducing network communications and 
increasing memory and CPU usage. As 
discussed in~\cite{zordan2014performance}, linear compression methods 
provide a very good compromise between these two factors, leading to 
substantial energy reduction.

% lossy vs lossless compression

% LTC summary
% define transmitted, received points, copmression ratio
The Lightweight Temporal Compression method 
(LTC~\cite{schoellhammer2004lightweight}) has been designed 
specifically for energy-constrained systems, initially sensor networks. 
It approximates data points by a piece-wise linear function that 
guarantees an upper bound on the reconstruction error, and a reduced 
memory footprint in $\mathcal{O}(1)$. However, LTC has only been 
described for 1D streams, while streams acquired by connected objects, such as 
acceleration or gyroscopic data, are often multi-dimensional. 

In this paper, we extend LTC to dimension $n$. To do so, we propose an 
algebraic formulation of the algorithm that also yields a 
norm-independent expression of it. We implement our extension on 
Motsai's Neblina module\footnote{\url{ https://motsai.com/products/neblina}}, and we test it on 3D acceleration streams 
acquired during human exercises, namely biceps curling, walking and 
running. Our implementation of LTC is available as free software.

We assume that the stream consists of a sequence of data points 
received at uneven intervals. The compression algorithm 
\emph{transmits} fewer points than it receives. The transmitted points 
might be included in the stream, or computed from stream points. The 
\emph{compression ratio} is the ratio between the number of received 
points and the number of transmitted points. An application 
reconstructs the stream from the transmitted points: the 
\emph{reconstruction error} is the maximum absolute difference between 
a point of the reconstructed stream, and the corresponding 
point in the original stream.

\setlength{\textfloatsep}{0pt}
\begin{figure}[b]
\centering
\includegraphics[width=\columnwidth]{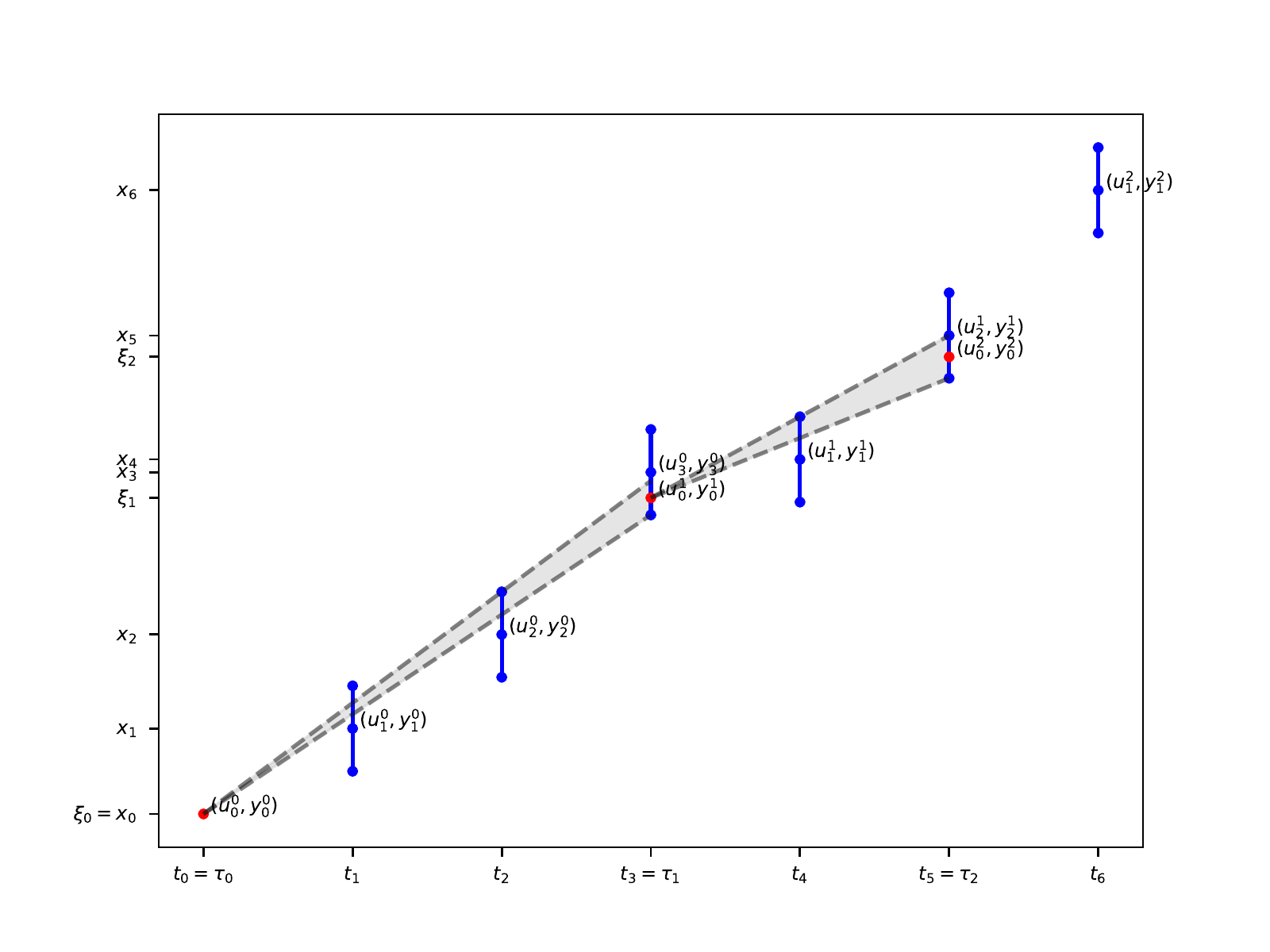}
\caption{Illustration of the LTC algorithm. Blue 
dots are received points, red dots are transmitted points. Dashed lines 
represent the high and low lines when a point is 
transmitted.\vspace*{-0.3cm}}
\label{fig:ltc}
\end{figure}

% Outline
Section~\ref{sec:ltc} provides some background on the LTC algorithm, 
and formalizes the description initially proposed 
in~\cite{schoellhammer2004lightweight}. Section~\ref{sec:extension} 
presents our norm-independent extension to dimension $n$, and 
Section~\ref{sec:implementation} describes our implementation. 
Section~\ref{sec:results} reports on experiments to validate our 
implementation, and evaluates the impact of n-dimensional LTC on 
energy consumption of connected objects.

\newpage

\section{Lightweight Temporal Compression}
\label{sec:ltc}

LTC approximates the data stream
by a piece-wise linear function of time, with an error bounded by parameter $\epsilon$.

\subsection{Notations}

The algorithm receives a stream of data points $x_i$
at times $t_i$ ($i \in \mathbb{N}$), and it transmits a stream of data points $\xi_i$
at times $\tau_i$ ($i \in \mathbb{N}$). To simplify the notations, we assume that:
\begin{equation*}
\forall k \in \mathbb{N}, \  \exists ! i \in \mathbb{N} \  \tau_k = t_i
\end{equation*}
That is, transmission times coincide with reception times.
We define the \emph{shifted received points} as follows:
\begin{equation*}
\forall k \in \mathbb{N}\ , \forall j \in \mathbb{N^*},\ (u^k_j, y^k_j) = (t_{i+j}, x_{i+j}), 
\end{equation*}
where $i$ is such that $t_i = \tau_k$ and:
\begin{equation*}
\forall k \in \mathbb{N},\  (u^k_0, y^k_0) = (\tau_k, \xi_k).
\end{equation*}
This definition is such that $y^k_j$ is the $j^{th}$ data point received
after the $k^{th}$ transmission and $u^k_j$ is the corresponding timestamp.
Figure~\ref{fig:ltc} illustrates the notations and algorithm.

The LTC algorithm maintains two lines, the \emph{high line}, and the
\emph{low line} defined by (1) the latest transmitted point and (2) the
\emph{high point} (high line) and the \emph{low point} (low line). When
a point ($t_i$, $x_i$) is received, the high line is updated as
follows: if $x_i+\epsilon$ is below the high line then the high line is
updated to the line defined by the last transmitted point and ($t_i$,
$x_i+\epsilon$); otherwise, the high line is not updated. Likewise, the low line
is updated from $x_i-\epsilon$. Therefore, any line located between the
high line and the low line approximates the data points received since
the last transmitted point with an error bounded by $\epsilon$.

\begin{algorithm}
\begin{algorithmic}[1]
\Input
   \Desc{$(u^k_j, y^k_j)$}{$\quad \quad $Received data stream}
   \Desc{$\epsilon$}{$\quad \quad$Error bound}
\EndInput
\Output
   \Desc{tr}{Transmitted points}
\EndOutput
\State tr = $(u^0_0, y^0_0)$ \Comment{Last transmitted point}
\State k = 0 ; j = 1
\State (lp, hp) = ($y^0_1 - \epsilon$, $y^0_1 + \epsilon$) \Comment{Low and high points}

\While{True} \Comment{Process received points as they come}
    \State j += 1
    \State new\_lp = max($y^k_j-\epsilon$, line($u^k_j$, tr, ($u^k_{j-1}$, lp)))
    \State new\_hp = min($y^k_j+\epsilon$, line($u^k_j$, tr, ($u^k_{j-1}$, hp)))
    \If{new\_lp $\leq$ new\_hp} \Comment{Keep compressing}
        \State (lp, hp) = (new\_lp, new\_hp)
    \Else
        \State tr = $(u^k_{j-1}, (lp+hp)/2)$
        \Comment{Transmit point}
        \State k += 1
        \State j = 1
        \State (lp, hp) = ($y^k_j-\epsilon$, $y^k_j+\epsilon$)
    \EndIf
\EndWhile
\end{algorithmic}
\caption{Original LTC algorithm, adapted from~\cite{schoellhammer2004lightweight}.}
\label{algo:ltc}
\end{algorithm}

Using these notations, the original LTC algorithm can
be written as in Algorithm~\ref{algo:ltc}. For readability, we assume
that access to data points is blocking, i.e., the program will wait
until the points are available. We also assume that the content of
variable \texttt{tr} is transmitted after each assignment of this
variable. Function \texttt{line}, omitted for brevity, returns the
ordinate at abscissa $x$ (1st argument) of the line defined by the points
in its 2nd and 3rd arguments.

\section{Extension to dimension $n$}
\label{sec:extension}
In this section we provide a norm-independent formulation of LTC in
dimension $n$. By $n$ we refer to the dimension of the data points
$x_i$. To handle time, LTC actually operates in dimension
$n+1$.

\subsection{Preliminary comments}

We note that the formulation of LTC 
in~\cite{schoellhammer2004lightweight} relies on the intersection of 
\emph{convex cones} in dimension $n+1$. For $n=1$, it corresponds to 
the intersection of triangles, which can efficiently be computed by 
maintaining boundary lines, as detailed previously. In higher 
dimension, however, cone intersections are not so straightforward, due 
to the fact that the intersection between cones may not be a cone.

To address this issue, we formulate LTC as an intersection test between
\emph{balls} of dimension $n$, that is, segments for $n=1$, disks for
$n=2$, etc. Balls are defined from the \emph{norm} used in
the vector space of data points. For $n=1$, the choice of the norm does
not really matter, as all p-norms and the infinity norm are identical.
In dimension $n$, however, norm selection will be critical.

\subsection{Algebraic formulation of LTC}

\subsubsection{Definitions}

Let $(u_0^k, y_0^k) \in \mathbb{R}^{n+1}$ be the latest transmitted point. For convenience, all the subsequent points will be
expressed in the orthogonal space with origin $(\tau_k, \xi_k)$. We denote by $(v_j, z_j)_{j \in \llbracket 0, m \rrbracket}$ such points:
\begin{equation*}
\forall j \leq m,\  (v_j, z_j) = (u_j^k - \tau_k, y_j^k - \xi_k)
\end{equation*}
Let $\mathcal{B}_j$ be the ball of $\mathbb{R}^n$ of centre $\frac{v_1}{v_j}z_j$ and radius
$\frac{v_1}{v_j}\epsilon$:
\begin{equation*}
\mathcal{B}_j = \left\{ z \in \mathbb{R}^n,\  \norm{z-\frac{v_1}{v_j}z_j} \leq \frac{v_1}{v_j}\epsilon \right\}
\end{equation*}
Note that $v_1$ is defined as soon as one point is received after the last
transmission.

\subsubsection{LTC property}

We define the \emph{LTC property} as follows:
\begin{equation*}
\exists z \in \mathbb{R}^n, \ \forall j \in \llbracket 1, m \rrbracket, \norm{\frac{v_j}{v_1}z-z_j} \leq \epsilon.
\end{equation*}
The original LTC algorithm ensures that the LTC property is
verified between each transmission. Indeed, all the data points
$z$ such that $(v_1, z)$ is between the high line and the low line
verify the property. Line 13 in Algorithm~\ref{algo:ltc} guarantees that
such a point exists.

The LTC property can be re-written as follows:
\begin{equation*}
\exists z \in \mathbb{R}^n, \ \forall j \in \llbracket 1, m \rrbracket, \norm{z-\frac{v_1}{v_j}z_j} \leq \frac{v_1}{v_j}\epsilon
\end{equation*}
that is:
\begin{equation}
\bigcap_{j=1}^m \mathcal{B}_j \neq \O
\label{eq:ltc-property}
\end{equation}
Note that $(\mathcal{B}_j)_{j \in \llbracket 1, m \rrbracket}$ is a sequence
of balls of strictly decreasing radius, since $v_j > v_1$.

\subsection{Algorithm}

The LTC algorithm generalized to dimension $n$ tests that the LTC 
property in Equation~\ref{eq:ltc-property} is verified after each reception of a data 
point. It is written in Algorithm~\ref{algo:general-ltc}.
\begin{algorithm}
\begin{algorithmic}[1]
\Input
   \Desc{$(u^k_j, y^k_j)$}{$\quad \quad $Received data stream}
   \Desc{$\epsilon$}{$\quad \quad$Error bound}
\EndInput
\Output
   \Desc{tr}{Transmitted points}
\EndOutput

\State tr = ($\tau, \xi$) = ($u^0_0, y^0_0$) \Comment{Last transmitted point}
\State k = 0 ; j = 0
\While{True}
    \State j += 1
    \State ($v_j, z_j$) = ($u_j^k - \tau, y_j^k - \xi$)
    \If{$\bigcap_{l=1}^j{\mathcal{B}_l} = \O$}
        \State Pick $z$ in $\bigcap_{l=1}^{j-1}{\mathcal{B}_j}$ \Comment{Transmit point}
        \State tr = ($\tau$, $\xi$) = ($u^k_{j-1}, z$)
        \State k += 1
        \State j = 1
    \EndIf
\EndWhile
\end{algorithmic}
\caption{Generalized LTC.}
\label{algo:general-ltc}
\end{algorithm}

\subsection{Ball intersections}

Although Algorithm~\ref{algo:general-ltc} looks simple, one should not
overlook the fact that there is no good general algorithm to test
whether a set of balls intersect. The best general algorithm we could find
so far relies on Helly's theorem which is formulated as follows~\cite{helly1923mengen}:
\begin{theorem}
Let $\left\{ X_i \right\}_{i \in \llbracket 1, m \rrbracket}$ be a collection of convex subsets of $\mathbb{R}^n$. If the intersection of every $n+1$
subsets is non-empty, then the whole collection has an non-empty intersection.
\end{theorem}
\noindent This theorem leads to an algorithm of complexity ${m \choose n+1}$ which is
not usable in resource-constrained environments.

The only feasible algorithm that we found is norm-specific. It
maintains a representation of the intersection
$\bigcap_{j=1}^{m}{\mathcal{B}_j}$ which is updated at every iteration.
The intersection tests can then be done in constant time. However,
updating the representation of the intersection may be costly
depending on the norm used. For the infinity norm, the representation
is a rectangular cuboid which is straightforward to update by
intersection with an n-ball.
For the Euclidean norm, the representation is a volume with no particular property,
which is more costly to maintain.

\subsection{Effect of the norm}

As mentioned before, norm selection in $\mathbb{R}^n$ has a critical
impact on the compression error and ratio. To appreciate this effect,
let us compare the infinity norm and the
Euclidean norm in dimension 2. By comparing the unit disk to a
square of side 2, we obtain that the compression ratio of a random stream would
be $\frac{4}{\pi}$ times larger with the infinity norm than with the 
Euclidean norm. In 3D, this ratio would be $\frac{6}{\pi}$. Conversely, 
a compression error bounded by $\epsilon$ with the infinity norm 
corresponds to a compression error of $\frac{\epsilon}{\sqrt{n}}$ with 
the Euclidean norm. Unsurprisingly, the
infinity norm is more tolerant than the Euclidean norm.

It should also be noted that using the infinity norm in $\mathbb{R}^n$ 
boils down to the use of the 1D LTC algorithm independently in each 
dimension, since a data point will be transmitted as soon as the linear 
approximation doesn't hold in any of the dimensions. For the Euclidean 
norm, however, the multidimensional and multiple unidimensional 
versions are different: the multiple unidimensional version 
behaves as the infinity norm, but the multidimensional version is more 
stringent, leading to a reduced compression rate and error.

To choose between the multidimensional implementation and multiple 
unidimensional ones, we recommend to check whether the desired 
error bound is expressed independently for every sensor, or as an aggregate error between them.
The multidimensional version is
more appropriate for multidimensional sensors, for instance 3D 
accelerometers or 3D gyroscopes, and the multiple unidimensional 
version is more suitable for multiple independent sensors, for 
instance a temperature and a pressure sensor.
%~ For instance, in case of a 3D accelerometer, if the error 
%~ is expressed simply as 
%~ ``10~mg", then the multidimensional version should be chosen because it 
%~ will guarantee that the norm of the 3D acceleration vector will be 
%~ reconstructed with an error less than 10~mg. Conversely, if the error is 
%~ specified as ``10~mg in x, 
%~ 10~mg in y and 10~mg in z", then the multiple unidimensional version 
%~ should be used.

\section{Implementation}
\label{sec:implementation}

To implement LTC in nD with the infinity norm, we maintain a cuboid 
representation of $\cap_{l=1}^j{\mathcal{B}_l}$ across the 
iterations of the \texttt{while} loop in 
Algorithm~\ref{algo:general-ltc}. The implementation works with
constant memory and requires limited CPU time.

With the Euclidean norm, the intersection test is more complex. We keep 
in memory a growing set $S$ of balls and the bounding box $B$ of their 
intersection. Then, when a new point arrives, we consider the 
associated ball $\mathcal{B}_j$ and our intersection test works as in 
Algorithm~\ref{algo:euclidean}. \texttt{box} is a function that returns 
the bounding box of an n-ball. \texttt{find\_bisection(S, B)} searches 
for a point in all the elements in S, using plane sweep and bisection 
initialized by the bounds of B. Our code is available at 
\url{https://github.com/big-data-lab-team/stream-summarization} under 
MIT license.

\begin{algorithm}
\begin{algorithmic}[1]
\Input
   \Desc{$S$}{Set of intersecting balls}
   \Desc{$B$}{Bounding box of the intersection of balls in $S$}
   \Desc{$\mathcal{B}_j$}{New ball to check}
\EndInput
\Output
   \Desc{$S$}{Updated set of intersecting balls}
   \Desc{$B$}{Updated bounding box}
   \Desc{$T$}{True if all the balls in S and $\mathcal{B}_j$ intersect}
\EndOutput

\If{$\mathcal{B}_j \cap B = \O$} \Comment{Ball is outside bounding box}
\State \Return ($S$, $B$, False) 
\EndIf
\If{$\exists\ \mathcal{B}_i \in S$ s.t. $\mathcal{B}_j \cap \mathcal{B}_i = \O$}
\State \Return ($S$, False) \Comment{Some balls don't intersect}
\EndIf
\If{$\exists\ \mathcal{B}_i \in S$ s.t. $\mathcal{B}_j \subset \mathcal{B}_i$} \Comment{Remove inclusions}
\State Remove $\mathcal{B}_i$ from $B$. Add $\mathcal{B}_j$ to $B$.
\EndIf
\State $B$ = box($\mathcal{B}_j$) $\ \bigcap\ $ $B$
\State $S$ = $S \ \bigcup \  \{\mathcal{B}_j\}$
\State $x$ = find\_bisection($S$, $B$) \Comment{This can take some time}
\If{$x$ == Null}
\State \Return ($S$, $B$, False)
\Else
\State \Return ($S$, $B$, True)
\EndIf
\end{algorithmic}
\caption{Intersection test for Euclidean n-balls.}
\label{algo:euclidean}
\end{algorithm}

\section{Experiments and Results}
\label{sec:results}

We conducted two experiments using Motsai's Neblina 
module, a system 
with a Nordic Semiconductor nRF52832 micro-controller, 64~KB of RAM, 
and Bluetooth Low Energy connectivity. Neblina has a 3D 
accelerometer, a 3D gyroscope, a 3D magnetometer, and environmental 
sensors for humidity, temperature and pressure. The platform is 
equipped with sensor fusion algorithms for 3D orientation tracking and 
a machine learning engine for complex motion analysis and motion 
pattern recognition~\cite{sarbishei2016accuracy}. Neblina has a 
battery of 100mAh; at 200~Hz, its average consumption is 2.52~mA when using 
accelerometer and gyroscope sensors but without radio 
transmission, and 3.47~mA with radio transmission, leading to an 
autonomy of 39.7~h without transmission and 28.8~h with transmission. 

\subsection{Experiment 1: validation}

%\todo{(Put a picture of Neblina in a box worn 
%on the wrist)}
\begin{wrapfigure}{R}{3cm}
\includegraphics[width=3cm]{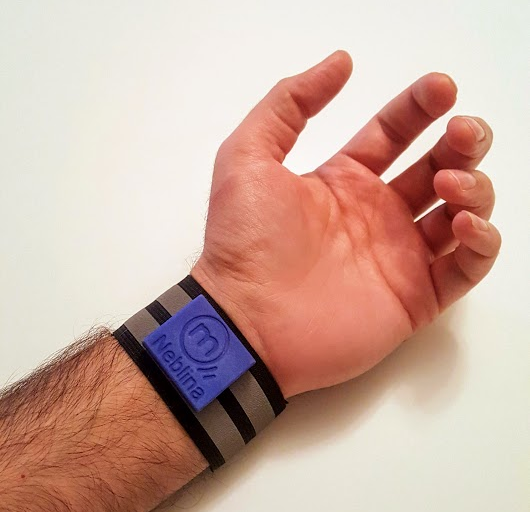}
\caption{Setup with Neblina.}
\label{fig:neblina-wrist}
\end{wrapfigure}
\begin{figure*}
\centering
\begin{subfigure}{1.85\columnwidth}
\centering
\includegraphics[width=0.3\columnwidth]{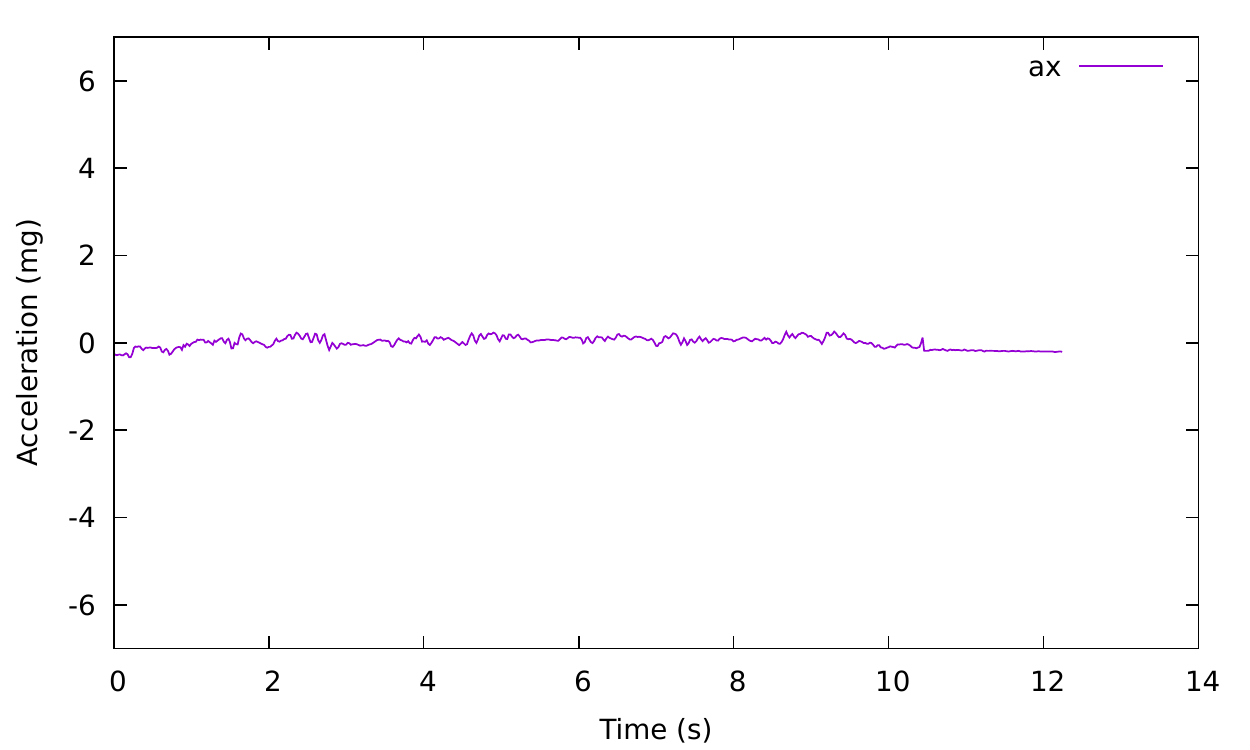}
\includegraphics[width=0.3\columnwidth]{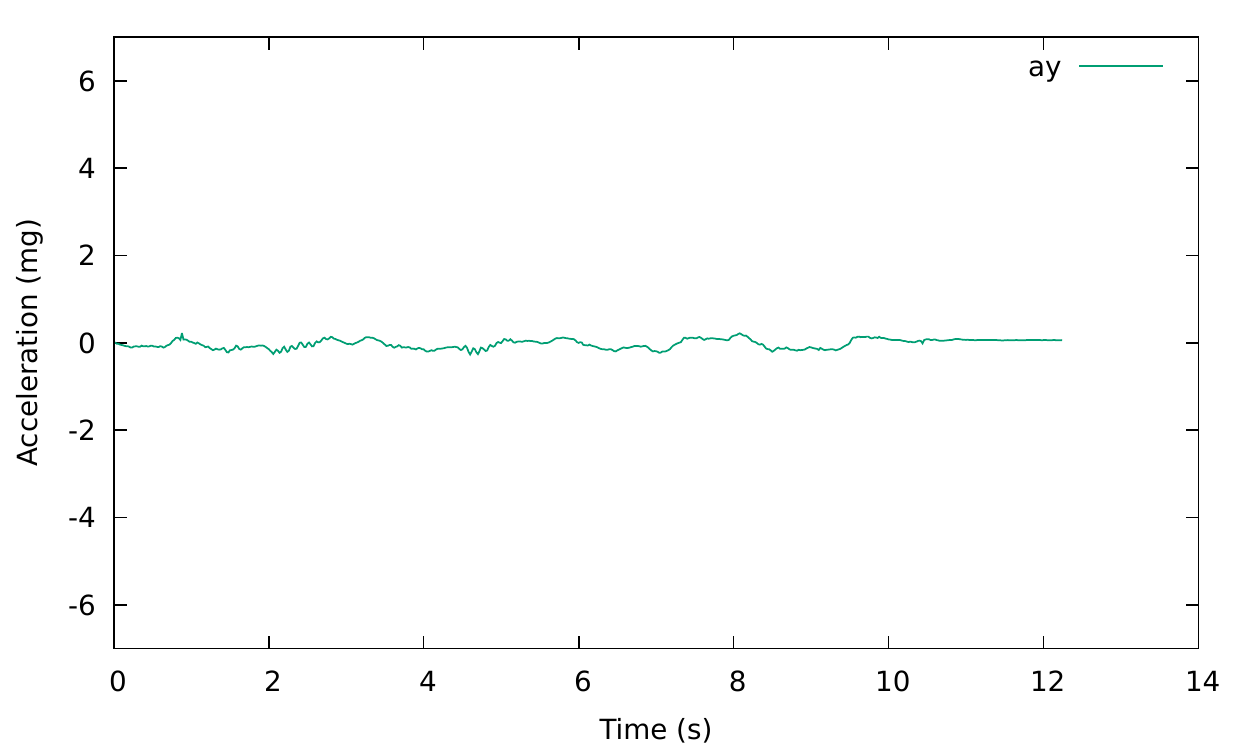}
\includegraphics[width=0.3\columnwidth]{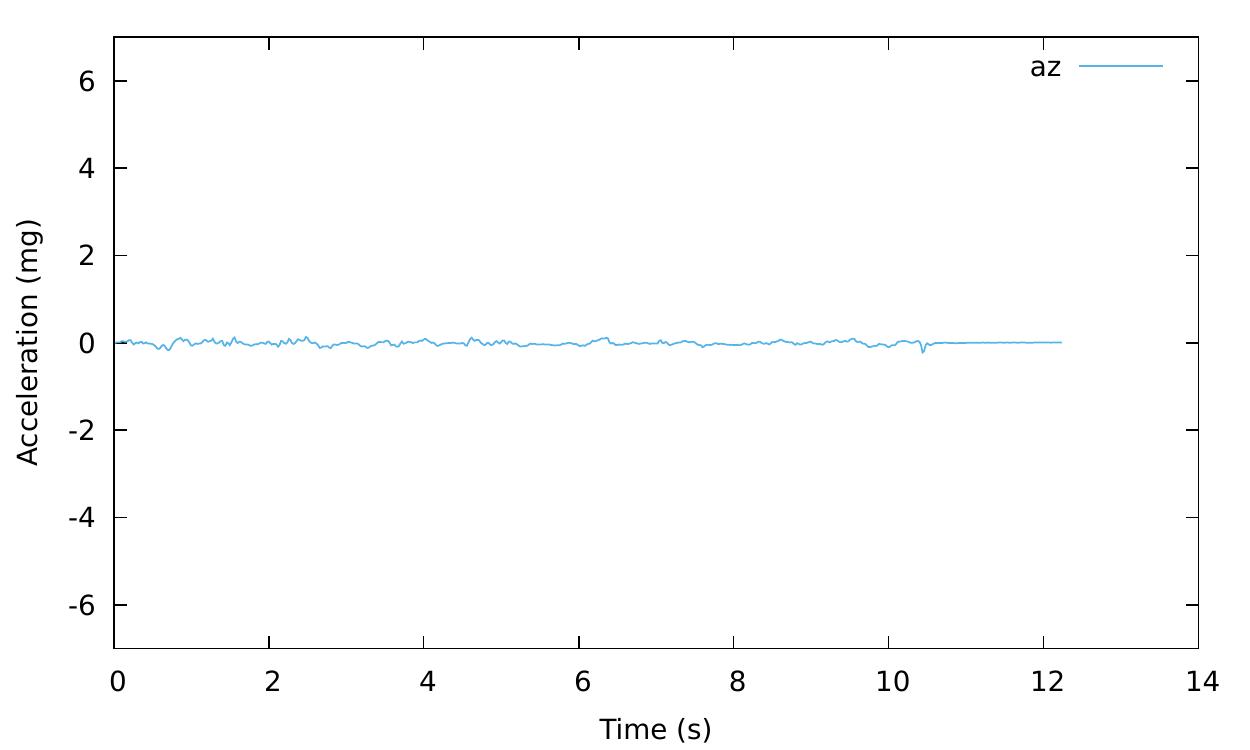}
\caption{Short biceps curl$^a$}
\end{subfigure}

{\footnotesize $^a$ Average 
of az data is -7.81mg. It was shifted to 0 so that the graphs can all 
use the same y scale.}

\begin{subfigure}{1.85\columnwidth}
\centering
\includegraphics[width=0.3\columnwidth]{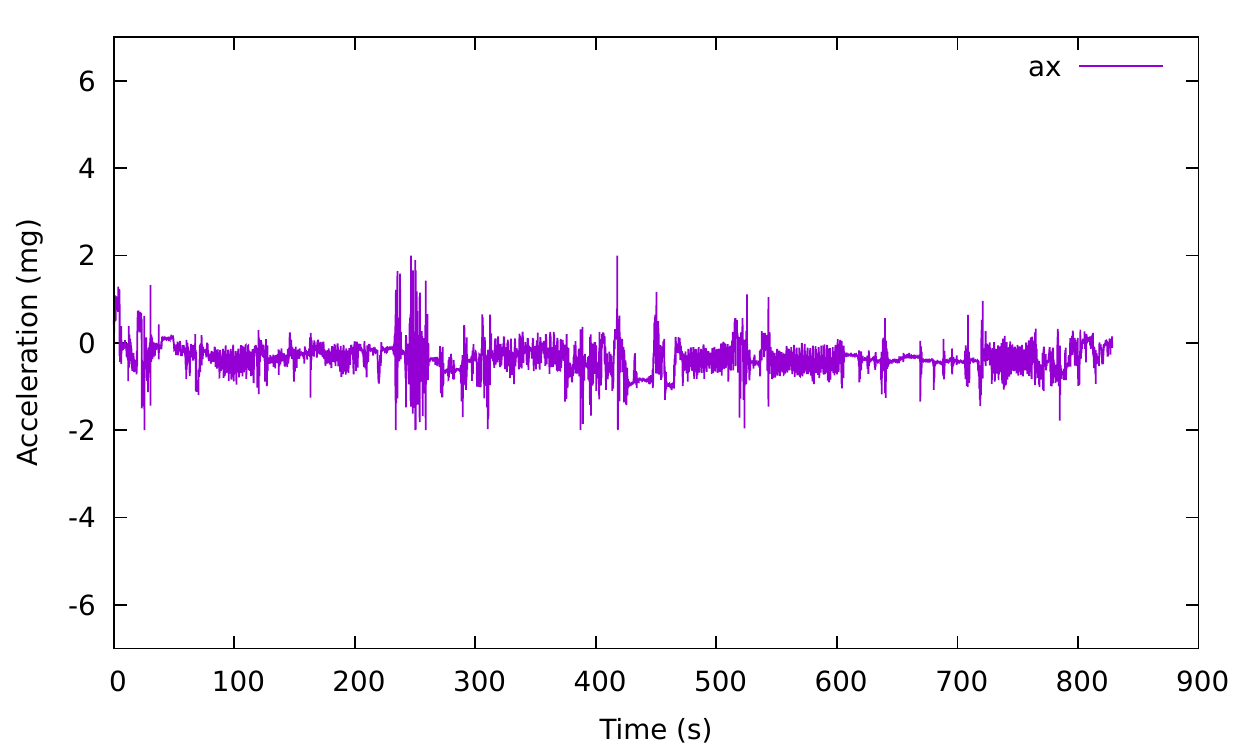}
\includegraphics[width=0.3\columnwidth]{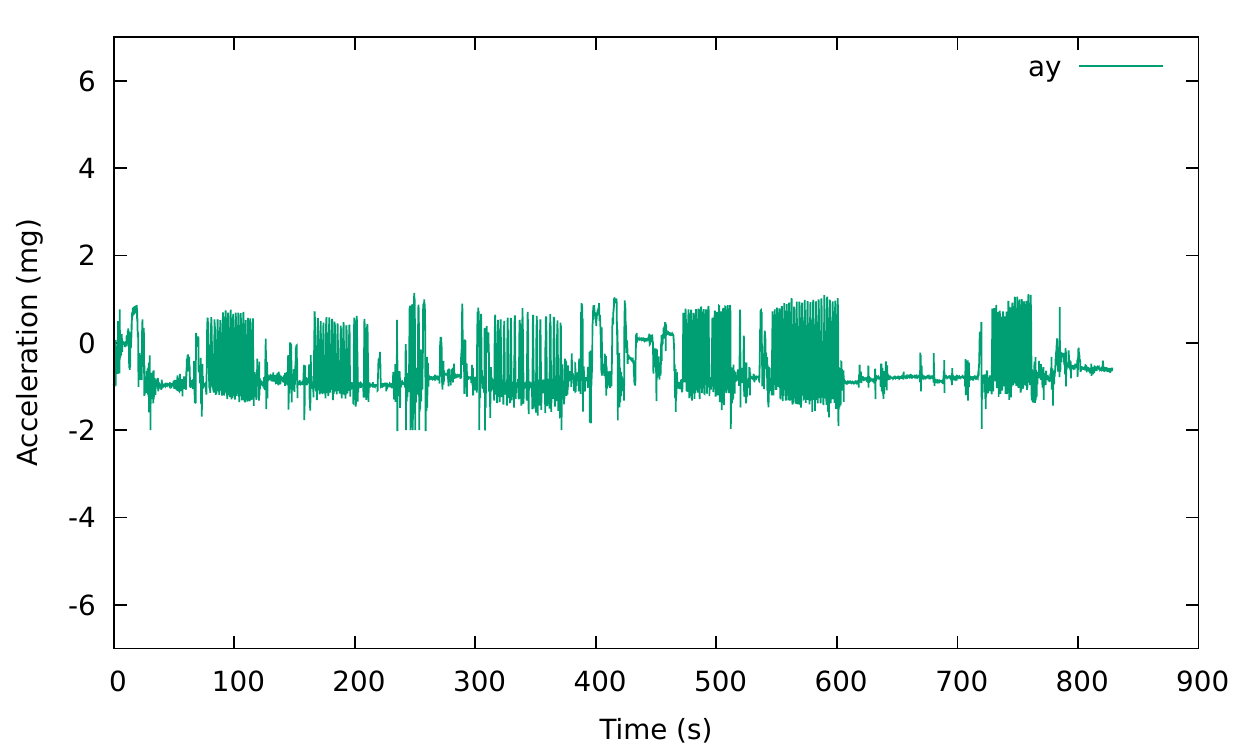}
\includegraphics[width=0.3\columnwidth]{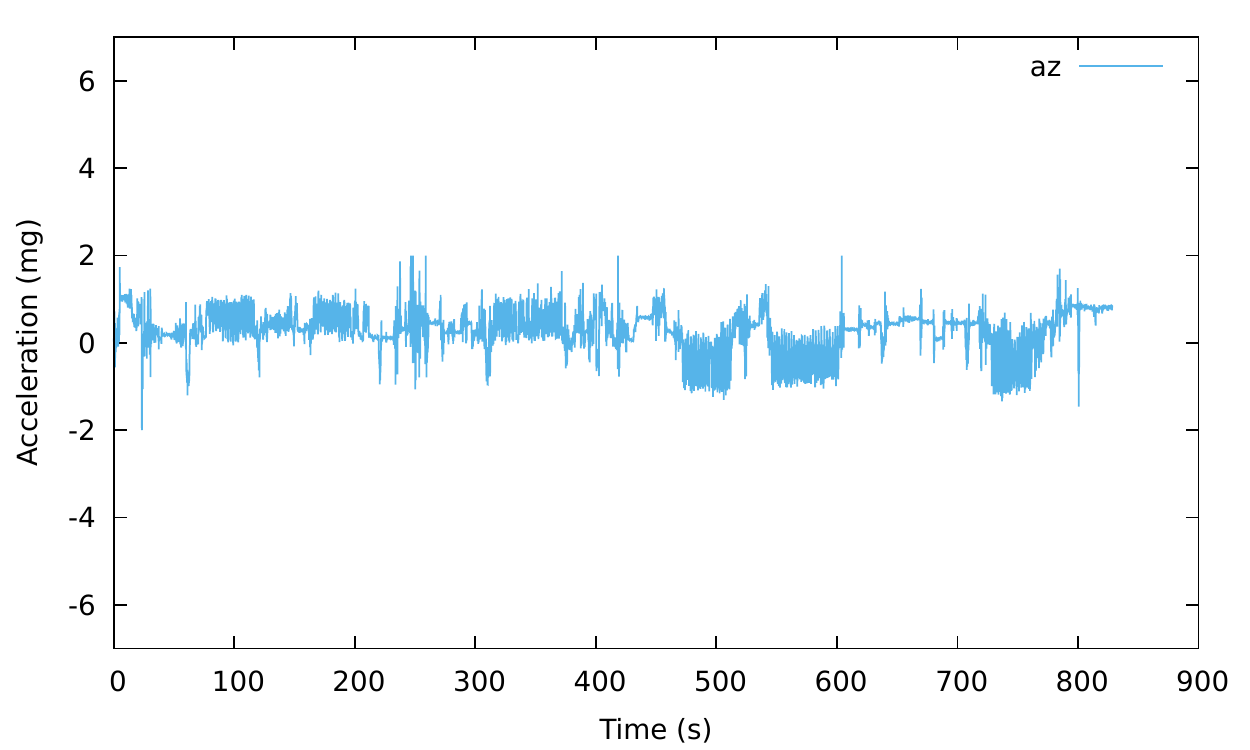}
\caption{Long biceps curl}
\end{subfigure}

\caption{Time-series used in Experiment 1}
\label{fig:datasets-1}
\end{figure*}

We validated the behaviour of our LTC extension on a PC using data 
acquired with Neblina. We collected two 3D accelerometer time-series, a 
short one and a longer one, acquired on two different subjects 
performing biceps curl, with a 50~Hz sampling rate (see 
Figure~\ref{fig:datasets-1}). In both cases, the subject was wearing 
Neblina on their wrist, as in Figure~\ref{fig:neblina-wrist}. It should be noted that the longest
time-series also has a higher amplitude, perhaps due to differences between 
subjects.

We compressed the time-series with various values of $\epsilon$, using 
our 2D (x and y) and 3D (x, y and z) implementations of LTC. On 
Neblina, the raw uncalibrated accelerometer data corresponds to errors 
around 20~mg (1~g is 9.8~m/s$^2$). We used a 
laptop computer with 16~GB of RAM, an Intel i5-3210M CPU @ 2.50GHz 
$\times$ 4, and Linux Fedora 27. We measured memory consumption using 
Valgrind's massif 
tool~\cite{nethercote2006building}, 
and processing time using \texttt{gettimeofday()} from the GNU C 
Library. 

Results are reported in Table~\ref{table:results-validation}. 
As expected, the compression ratio increases with $\epsilon$, and the 
maximum measured error remains lower than $\epsilon$ in all cases. The 
maximum is reached most of the time on these time-series.

\paragraph{Infinity vs Euclidean norms}
The average ratio between the compression ratios obtained
with the infinity and Euclidean norm is 1.03 for 2D data, and 1.06
for 3D data. These ratios are lower than the theoretical values of
$\frac{4}{\pi}$ in 2D and $\frac{6}{\pi}$ in 3D, which are obtained for
random-uniform signals. Unsurprisingly, the infinity norm surpasses the
Euclidean norm in terms of resource consumption. Memory-wise, the
infinity norm requires a constant amount of 80~B, used to store the
intersection of n-balls. The Euclidean norm, however, uses up to 4.7~KB of memory
for the Long time-series in 3D with $\epsilon$=48.8~mg. More importantly,
the amount of required memory increases for longer time-series, and it also increases with larger
values of $\epsilon$. Similar observations are made for the processing
time, with values ranging from 0.4~ms for the simplest time-series and
smallest $\epsilon$, to 41.3~ms for the most complex time-series and
largest~$\epsilon$.
%~ Figure~\ref{fig:memory} shows the memory consumption
%~ of the 3D Euclidean implementation for $\epsilon$=48.8~mg: peaks appear
%~ at the end of long compressed ranges where the signal was $\epsilon$-closed
%~ to the linear approximation.

\paragraph{2D vs 3D}
For a given $\epsilon$, the compression
ratios are always higher in 2D than in 3D. It makes sense since the
probability for the signal to deviate from a straight line
approximation is higher in 3D than it is in 2D. Besides, resource
consumption is higher in 3D than in 2D: for the infinity norm, 3D
consumes 1.4 times more memory than 2D (1.8 times on average for
Euclidean norm), and processing time is 1.35 longer (1.34 on
average for Euclidean norm).

\begin{table}
    \begin{subfigure}{\columnwidth}
    \centering
    \begin{tabular}{l|l|l|l|l}
    \hline
    \rowcolor{headcolor}
                           & \multicolumn{2}{c|}{Infinity} & \multicolumn{2}{c}{Euclidean}\\
    \rowcolor{headcolor}
    $\epsilon$  (mg)          & 48.8         & 34.5       & 48.8       & 34.5 \\
    \hline
    Max error   (mg)          & 48.8         & 34.4       & 48.8       & 34.5 \\
    Compression ratio (\%)    & 79.77        & 72.59      & 77.49      & 70.96\\
    Peak memory   (B)         & 80           & 80         & 688        & 688  \\
    Processing time (ms)      & 0.101        & 0.094      & 0.456      & 0.406\\ \hline
    \end{tabular}
    \caption{Short biceps curl (2D)}
    \end{subfigure}\\
    \begin{subfigure}{\columnwidth}
    \centering
    \begin{tabular}{l|l|l|l|l}
    \hline
    \rowcolor{headcolor}
                   & \multicolumn{2}{c|}{Infinity} & \multicolumn{2}{c}{Euclidean} \\
    \rowcolor{headcolor}
    $\epsilon$ (mg)            & 48.8        & 34.5       & 48.8        & 34.5    \\
    \hline
    Max error  (mg)            & 48.8        & 34.5       & 48.8        & 34.5             \\
    Compression ratio (\%)     & 77.46       & 70.98      & 75.77       & 68.81           \\
    Peak memory  (B)           & 80          & 80         & 2512        & 2608             \\
    Processing time (ms)       & 6.06        & 5.84       & 33.84       & 31.07           \\ \hline
    \end{tabular}
    \caption{Long biceps curl (2D)}
    \end{subfigure}\\
    \begin{subfigure}{\columnwidth}
    \centering
    \begin{tabular}{l|l|l|l|l}
    \hline
    \rowcolor{headcolor}
                           & \multicolumn{2}{c|}{Infinity} & \multicolumn{2}{c}{Euclidean} \\
    \rowcolor{headcolor}
    $\epsilon$ (mg)        & 48.8          & 28.2          & 48.8   & 28.2   \\
    \hline
    Max error  (mg)        & 48.8          & 28.2          & 48.8   & 28.2   \\
    Compression ratio (\%) & 78.14         & 66.39         & 74.39  & 63.13   \\
    Peak memory (B)        & 112           & 112           & 1744   & 784    \\
    Processing time (ms)   & 0.147         & 0.134         & 0.731  & 0.514  \\ \hline
    \end{tabular}
    \caption{Short biceps curl (3D)}
    \end{subfigure}\\
    \begin{subfigure}{\columnwidth}
    \centering
    \begin{tabular}{l|l|l|l|l}
    \hline
    \rowcolor{headcolor}
                              & \multicolumn{2}{c|}{Infinity} & \multicolumn{2}{c}{Euclidean} \\
    \rowcolor{headcolor}
    $\epsilon$ (mg)                & 48.8        & 28.2       & 48.8     & 28.2    \\
    \hline
    Max error  (mg)                & 48.8        & 28.2       & 48.8     & 28.2    \\
    Compression ratio (\%)         & 71.23       & 58.11      & 67.35    & 53.24   \\
    Peak memory (B)                & 112         & 112        & 4752     & 3856    \\
    Processing time (ms)           & 7.87        & 7.22       & 41.29    & 39.04   \\ \hline
    \end{tabular}
    \caption{Long biceps curl (3D)}
    \end{subfigure}
    \caption{Results from Experiment 1}
    \label{table:results-validation}
\end{table}

\begin{figure*}
\centering
\begin{subfigure}{1.85\columnwidth}
\centering
\includegraphics[width=0.3\columnwidth]{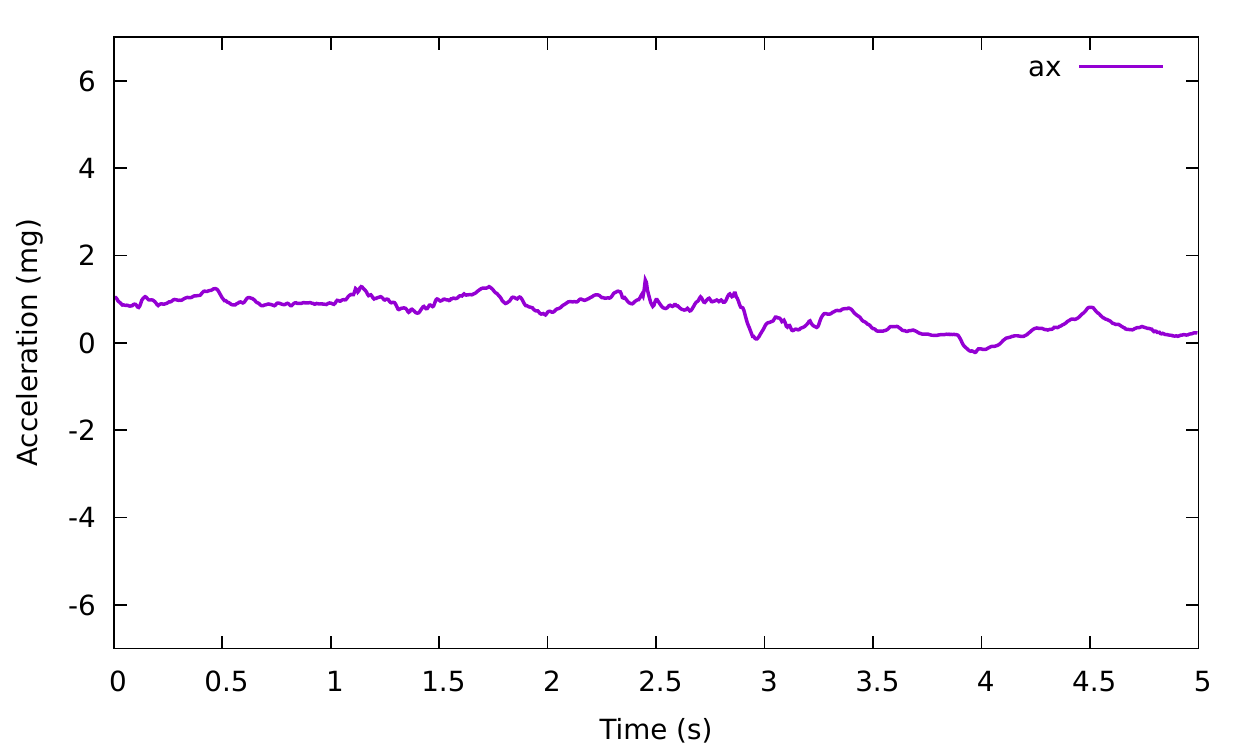}
\includegraphics[width=0.3\columnwidth]{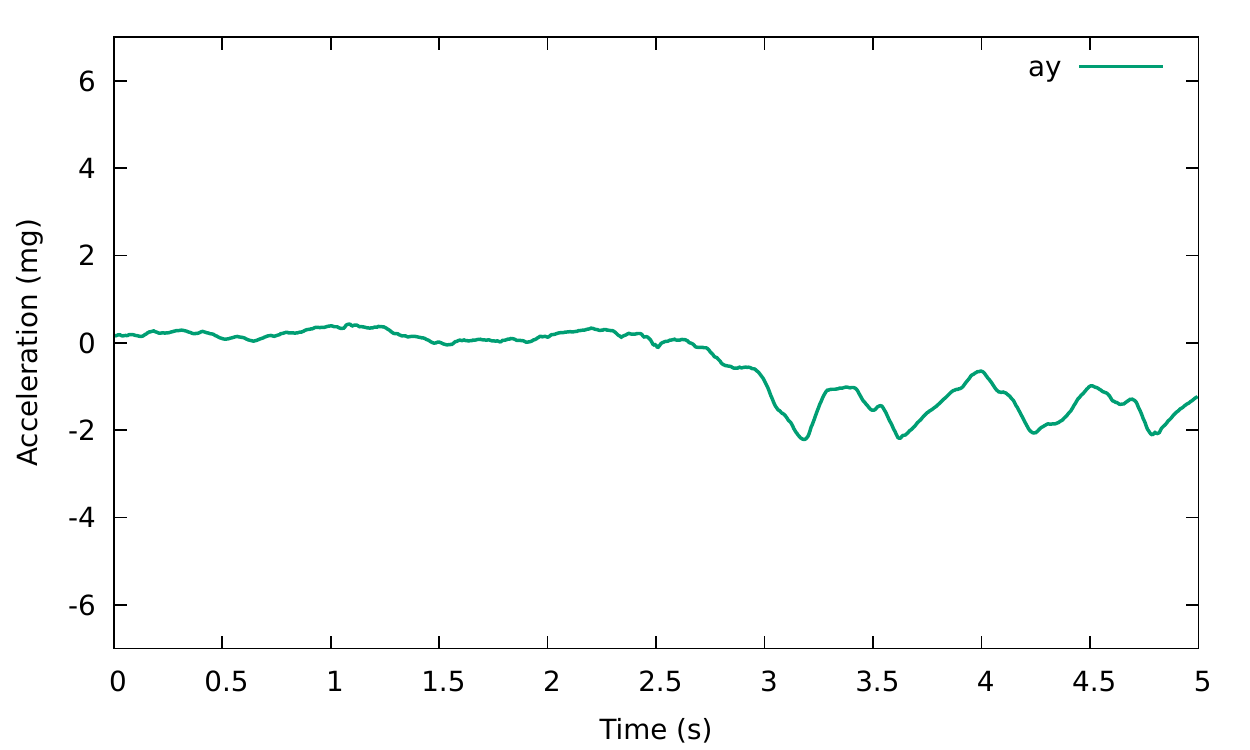}
\includegraphics[width=0.3\columnwidth]{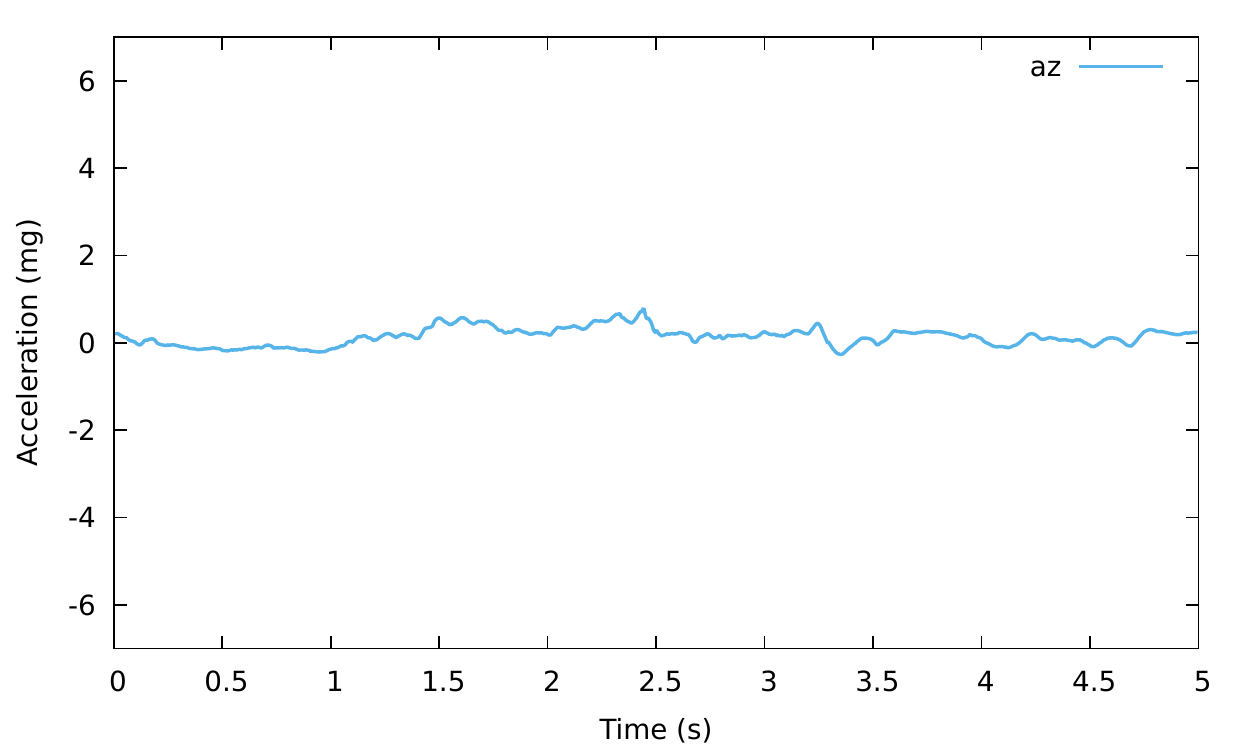}
\caption{Walking}
\end{subfigure}
\begin{subfigure}{1.85\columnwidth}
\centering
\includegraphics[width=0.3\columnwidth]{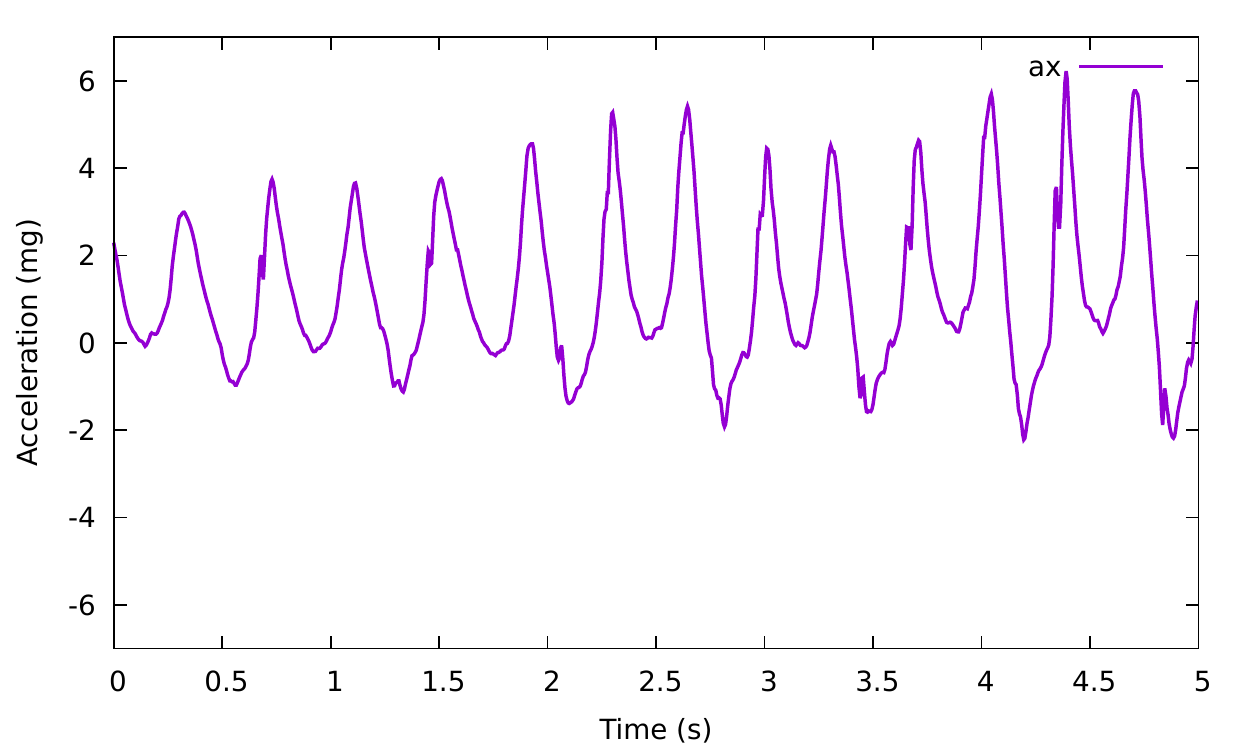}
\includegraphics[width=0.3\columnwidth]{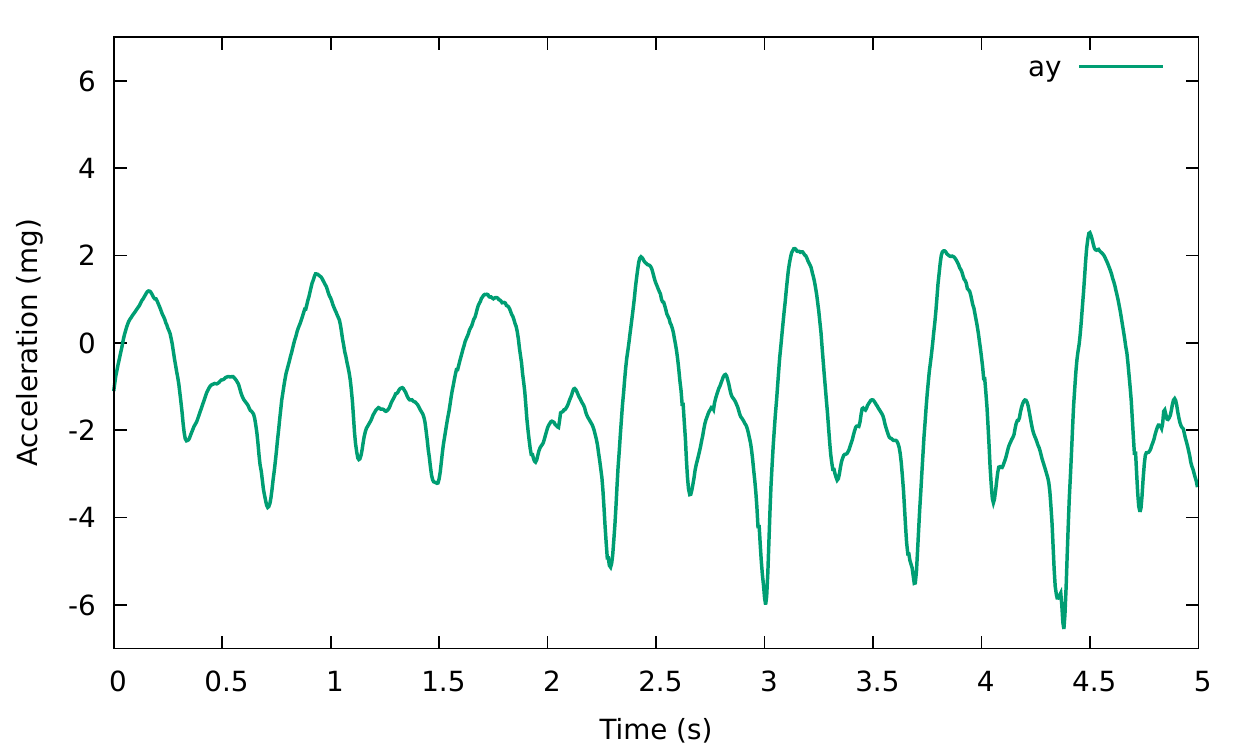}
\includegraphics[width=0.3\columnwidth]{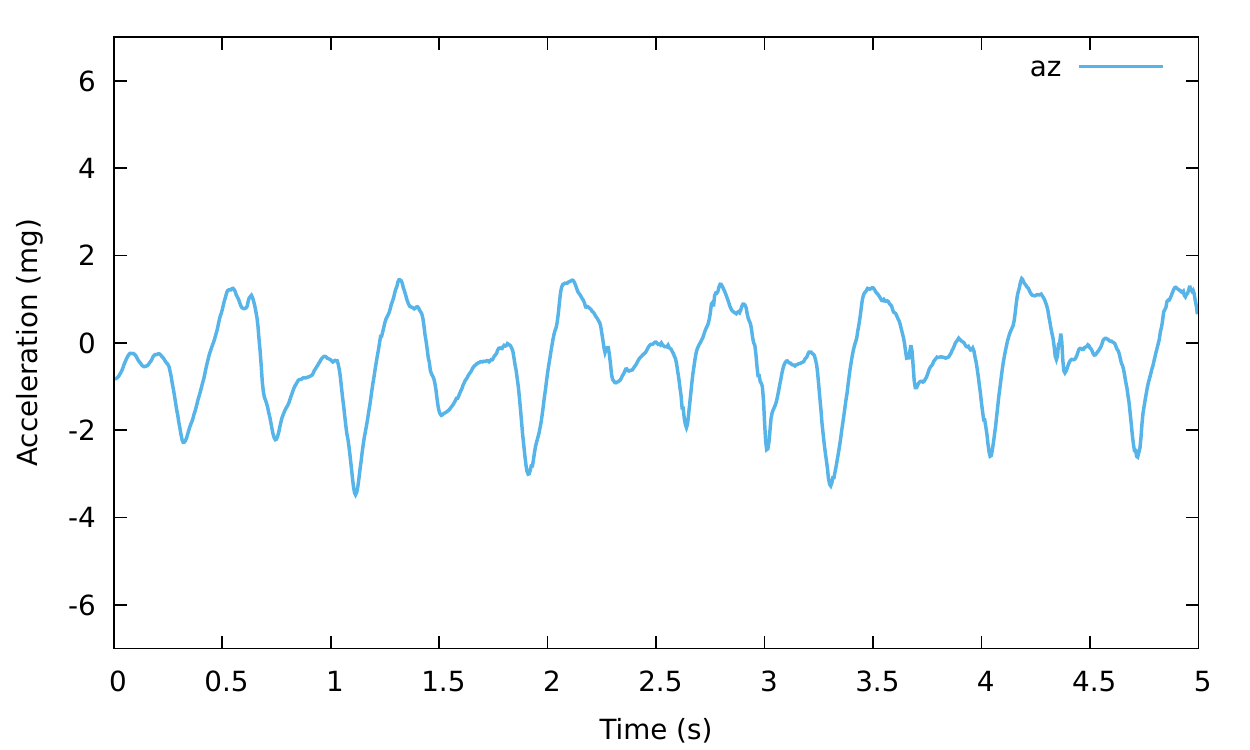}
\caption{Running}
\end{subfigure}
\caption{Time series used in Experiment 2 \vspace*{-0.3cm}}
\label{fig:datasets-2}
\end{figure*}

%~ \begin{figure}
%~ \includegraphics[width=\columnwidth]{./figures/memory-mohammad.pdf}
%~ \caption{Memory usage of 3D Euclidean implementation, long biceps curl time-series, $\epsilon$=48.8~mg.}
%~ \label{fig:memory}
%~ \end{figure}

\subsection{Experiment 2: impact on energy consumption}

We acquired two 3D accelerometer time-series at 200~Hz for two 
activities: walking and running (see Figure~\ref{fig:datasets-2}). In 
both cases, the subject was wearing Neblina on their wrist as in 
Experiment 1. We collected 1,000 data points for each activity, 
corresponding to 5 seconds of activity.

We measured energy consumption associated with the transmission of 
these time-series by ``replaying" the time-series after loading them as 
a byte array in Neblina. We measured the current every 500~ms. We also 
measured the max and average latency resulting from compression.

 Results are reported in Table~\ref{table:results-energy}. For a given
 $\epsilon$ and norm, the compression ratio is larger for walking than
 for running. The ratio of saved energy is relative to the reference
 current of 3.47~mA measured when Neblina transmits data without
 compression. In all cases, activating compression saves energy. The 
 reduction in energy consumption behaves as the compression ratio: it 
 increases with $\epsilon$, it is higher for the infinity norm than for 
 the Euclidean, and it is higher for the walking activity than for 
 running. For a realistic error of $\epsilon$=9.8~mg, the ratio of 
 saved energy with the infinity norm is close to 20\% for the walking 
 activity, which is substantial. Latency is higher for walking 
 than for running, and it is also higher for the Euclidean norm than 
 for the infinity norm. In all cases, the latency remains lower 
 than the 5-ms tolerable latency at 200~Hz, which demonstrates the 
 feasibility of 3D LTC compression.

\begin{table}[]
   \begin{subfigure}{\columnwidth}
   \centering
   \begin{tabular}{l|l|l|l|l|l|l}
   \hline
   \rowcolor{headcolor}
                          & \multicolumn{3}{c|}{Infinity} & \multicolumn{3}{c}{Euclidean} \\
   \rowcolor{headcolor}
   $\epsilon$ (mg)             & 48.8    & 9.8      & 4.9   & 48.8   & 9.8   & 4.9   \\
   \hline
   Max error (mg)              & 48.8    & 9.8      & 4.9   & 48.8   & 9.8   & 4.9   \\
   Compr.      ratio (\%)      & 88.9    & 66.4     & 45.5  & 87.6   & 63.3  & 37.2  \\
   Average (mA)                & 2.64    & 2.79     & 3.02  & 3.10   & 3.02  & 3.13  \\
   Saved energy (\%)           & 23.9    & 19.7     & 13.0  &  10.7  & 13.0  & 9.7   \\
   Max latency ($\mu$s)& 60      & --       & --    & 1530   & --    & --    \\
   Average latency ($\mu$s) & 31 & --       & --    & 145    & --    & --    \\    
   \hline
   \end{tabular}
   \caption{Walking}
   \end{subfigure}
   \begin{subfigure}{\columnwidth}
   \centering
   \begin{tabular}{l|l|l|l|l|l|l}
   \hline
   \rowcolor{headcolor}

                     & \multicolumn{3}{c|}{Infinity} & \multicolumn{3}{c}{Euclidean} \\
    \rowcolor{headcolor}
   $\epsilon$ (mg)        & 48.8       & 9.8      & 4.9       & 48.8      & 9.8    & 4.9    \\
   \hline
   Max error (mg)         & 48.8       & 9.8      & 4.9       & 48.8      & 9.8    & 4.9   \\
   Compr.      ratio (\%) & 68.6       & 25.5     & 9.5       & 64.4      & 19.8   & 5.7   \\
   Average (mA)        & 2.88     & 3.22   & 3.38     & 2.95    & 3.32    & 3.39   \\
   Saved energy (\%)      & 17.0    & 7.2    & 2.5   & 14.9   & 4.3   & 2.2\\
   Max latency ($\mu$s)& 60      & --       & --    & 840   & --    & --    \\
   Average latency ($\mu$s) & 30 & --       & --    & 64    & --    & --    \\    
   \hline
   \end{tabular}
   \caption{Running}
   \end{subfigure}
   \caption{Results from Experiment 2}
   \label{table:results-energy}
\end{table}

\section{Conclusion}

We presented an extension of the Lightweight Temporal Compression 
method to dimension $n$ that can be instantiated for any distance 
function. Our extension formulates LTC as an intersection detection 
problem between n-balls. We implemented our extension on Neblina for the 
infinity and Euclidean norms, and we measured the energy reduction 
induced by compression for acceleration streams acquired during
 human activities.

We conclude from our experiments that the proposed extension to LTC is 
well suited to reduce energy consumption in connected objects. The 
implementation behaves better with the infinity norm than with the 
Euclidean one, due to the time complexity of the current algorithm to 
detect the intersection between n-balls for the Euclidean norm. 

Our future work focuses on this latter issue. We plan to start from 
Helly's theorem, which only provides an algorithm of complexity ${m 
\choose n+1}$ to compress $m$ points in 
dimension $n$. We note that Helly's theorem holds for arbitrary convex 
subsets of $\mathbb{R}^n$, while we are considering a \emph{sequence} 
of \emph{balls} of \emph{decreasing radius}. Based on this 
observation, a stronger result might exist that would lead to a more 
efficient implementation of LTC with the Euclidean norm. Our current 
idea is to search for a point expressed as a function of the ball centres that would 
necessarily belong to the ball intersection when it is not empty; such 
a point, if it exists, necessarily converges to the centre of the last 
ball in the sequence as $n$ increases, as the radius of the last ball 
decreases to zero. The resulting algorithm would then compute this 
point and check its inclusion in every ball, which is done in 
$\mathcal{O}(m)$ complexity.

The choice of an appropriate norm should not be underestimated. Some 
situations might be better described with the Euclidean norm than with 
the infinity norm, such as the ones involving position or movement 
measures. Using the infinity norm instead of the Euclidean would lead 
to important error differences, proportional to $\sqrt{n}$ in dimension 
$n$. Investigating other norms, in 
particular the 1-norm, would be relevant too.

\section*{Acknowledgement}
This work was funded by the Natural 
Sciences and Engineering Research Council of Canada.

\bibliographystyle{IEEEtran}
\bibliography{IEEEabrv,biblio.bib}

\end{document}